\documentclass{jpsj2}

\def\runauthor{Katahira et al.}

\newcommand{\Prob}{\mbox{Prob}}
\newcommand{\sgn}{\mbox{sgn}}
\newcommand{\erf}{\mbox{erf}}
\renewcommand{\vec}[1]{\mbox{\boldmath $#1$}}

\title{%
Retrieval of branching sequences in an associative memory model with common external input and bias input 
}

\author{%
Kentaro \textsc{Katahira}\thanks{E-mail address: katahira@mns.k.u-tokyo.ac.jp}$^{1,2}$, 
Masaki \textsc{Kawamura}$^{3}$, Kazuo \textsc{Okanoya}$^{2}$, and Masato \textsc{Okada}$^{1,2}$
}

\inst{%
$^{1}$Graduate School of Frontier Sciences, 
The University of Tokyo\\ 5-1-5 Kashiwanoha, Kashiwa, 277-8561 Japan \\

$^{2}$RIKEN Brain Science Institute 2-1 Hirosawa, Wako, 351-0198 Japan

$^{3}$Graduate School of Science and Engineering, 
Yamaguchi University, \\1677-1 Yoshida, Yamaguchi 753-8512 Japan
}

\recdate{\today}

\abst{
We investigate a recurrent neural network model with common external and 
bias inputs that can retrieve branching sequences. Retrieval of memory 
sequences is one of the most important functions of the brain. A lot of 
research has been done on neural networks that process memory sequences. 
Most of it has focused on fixed memory sequences. However, many animals 
can remember and recall branching sequences. Therefore, we propose 
an associative memory model that can retrieve branching sequences. 
Our model has bias input and common external input. 
Kawamura and Okada reported that common external input 
enables sequential memory retrieval in an associative memory model 
with auto- and weak cross-correlation connections. 
We show that retrieval processes along branching 
sequences are controllable with both the bias input and the common 
external input. To analyze the behaviors of our model, we derived the 
macroscopic dynamical description as a probability density function. The 
results obtained by our theory agree with those obtained by computer 
simulations. 
}
\kword{%
branching sequence, common external input, bias input, associative 
memory, macroscopic dynamic description 
}

\begin{document} 
\maketitle

\section{Introduction} 
Retrieval of memory sequences is one of the most important functions of 
the brain. There have been many studies about neural networks that 
process memory sequences 
\cite{Amari1972,Sompolinsky1986,Gutfreund1988,Kitano1998,Katayama2001}, 
In these models, cross-correlation between one pattern and the next one 
in the sequence is embedded in synaptic connections. These connections 
make the model retrieves the sequential memory deterministically. The 
stored sequences are confined to fixed sequences without branching 
points. However, many animals, including humans, memorize and retrieve 
memory sequences with branching points. A specific example observed in 
songbirds, which have attracted interest as model animals 
for sequence learning and generation. 
Bengalese finches sing songs with finite state syntax 
that are composed of chunks of song notes and branching points of note sequences\cite{Okanoya2004}. Because such song patterns can be considered to be rudiments of human language syntax, and thus it is important to investigate the neural mechanisms for storing and retrieving branching sequences. Guyon et al. treat branching sequences with various synaptic transmission delays \cite{Guyon}. In their model, however, the state transitions are deterministic, depending on neural activity history. This makes it hard to dynamically select the next state at a branching point and to transit stochastically. 

Aoyagi and Aoki\cite{AoyagiAoki2004, AokiAoyagi2004} have investigated a model with auto- and weak cross-correlation connections and showed that transitions between attractors can be invoked by synchronous spikes. Kawamura and Okada assumed synchronous spikes to be correlated noises, i.e., common external input, and theoretically investigated the dynamics of associative memory models with common external input\cite{Kawamura2006}. In this paper, we propose an associative memory model with common external input and bias input that can store and retrieve branching sequences. Each memory pattern is stored as an attractor in an auto-correlation connection. Sequences among the attractors are stored in weak cross-correlation connections. Common external input can induce state transitions along the stored sequences. Since the distances among all memory patterns are equal, the model is trapped in a mixed state of memory patterns at a branching point. We introduce bias input to transit from the mixed state to the desired memory state. Due to common external input, the states of neurons are correlated. Therefore, the model has sample dependence\cite{Amari2003}. The behavior of such a model can be described by a probability density function of order parameters, i.e. overlap \cite{YamanaOkada2005, Kawamura2005, Kawamura2006}. We derive a probability density function (PDF) of macroscopic state variables and discuss the properties of our model. 

This paper is organized as follows. In the next section, we propose a model with common external input and bias input. In \S3, we discuss a model without bias input and present behavioral results for the model obtained theoretically and by computer simulations. In \S4, we discuss a model with bias input. We show that bias input enables our model to transit to a target memory pattern. The final section concludes the paper. \\

\section{Model} 
We consider an associative memory model consisting of $N$ neurons. The state of the $i$th neuron takes $x^t_i=\pm1$ at time $t$ and is updated synchronously by 
\begin{equation} 
x_i^{t+1} = F\left(\sum_{j=1}^NJ_{ij}x_j^t+\zeta_i^t+\eta^t + c B^t_i\right), 
\label{eqn:dynamics} 
\end{equation} 
where the output finction is $F(h)=\sgn(h)$, and $J_{ij}$ is a synaptic connection from the $j$th neuron to the $i$th neuron. $\zeta_i^t$ is independent external input for the $i$th neuron at time $t$. $\eta^t$ is common external input, which affects all neurons similarly. 

$B^t_i$ is also external input and is correlated with a particular memory pattern. Since this external input has an effect that biases the state toward a target pattern at a branching point in the sequences, we call it bias input. Bias input $B^t_i$ is generated by 
\begin{align} 
\Prob\left[B^t_i=\pm 1\right] = \frac{1\pm \sum^p_{\mu=1}b_t^{\mu}\xi_i^{\mu}}{2}, 
\label{eq:bias_input} 
\end{align} 
where 
\begin{align} 
\sum^p_{\mu=1}b_t^{\mu} \le 1, 
\end{align} 
and $b_t^{\mu}$ is overlap between bias input $B^t_i$ and the memory pattern $\vec{\xi}^\mu$ at time $t$. The constant, $c$, in (\ref{eqn:dynamics}) is the parameter that determines amplitude of the bias input. 

\begin{figure}[tb] 
\begin{minipage}{.45\linewidth} 
\begin{center} \includegraphics[width=.40\linewidth]{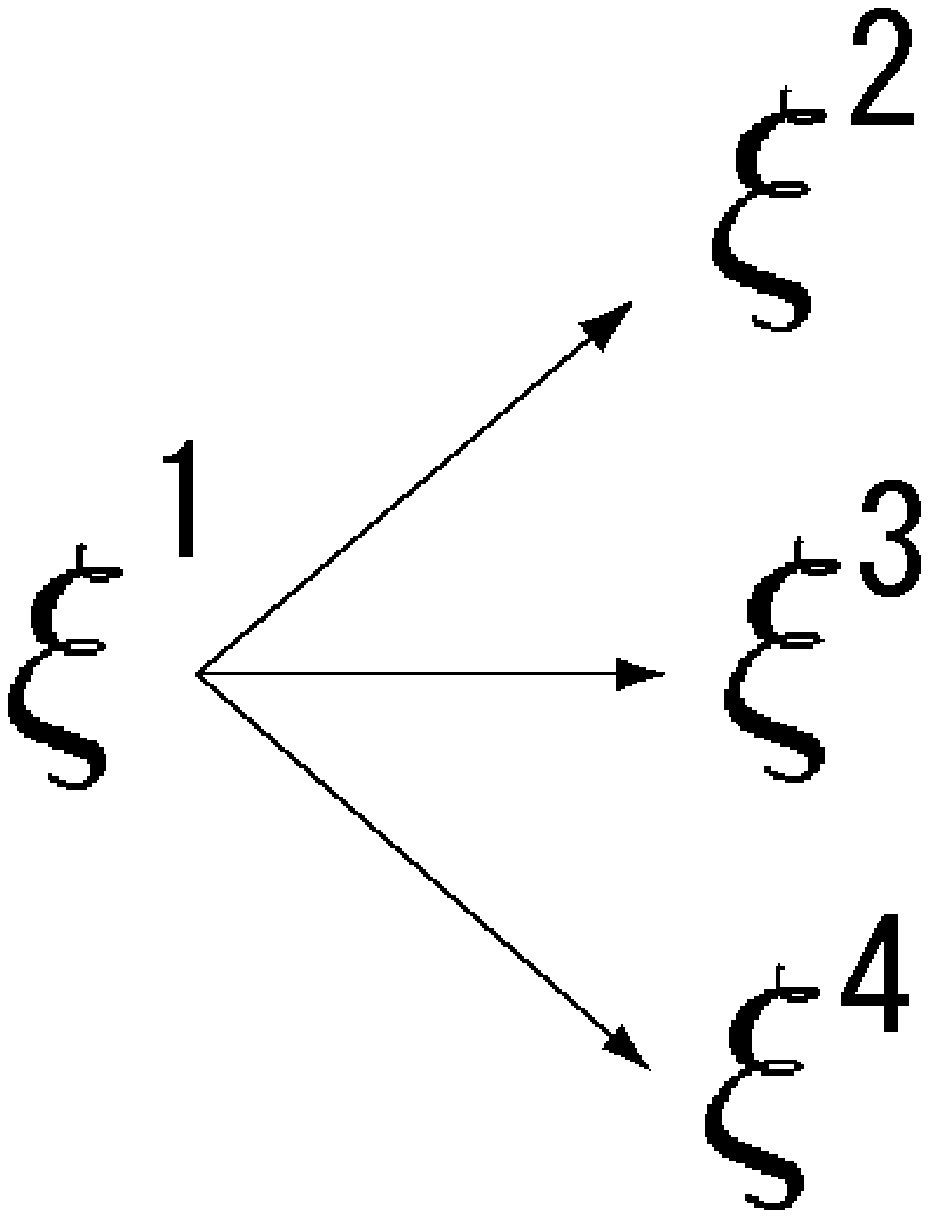} \\ (a) Sequence A. 
\end{center} 
\end{minipage} 
\begin{minipage}{.50\linewidth} 
\begin{center} \includegraphics[width=.75\linewidth]{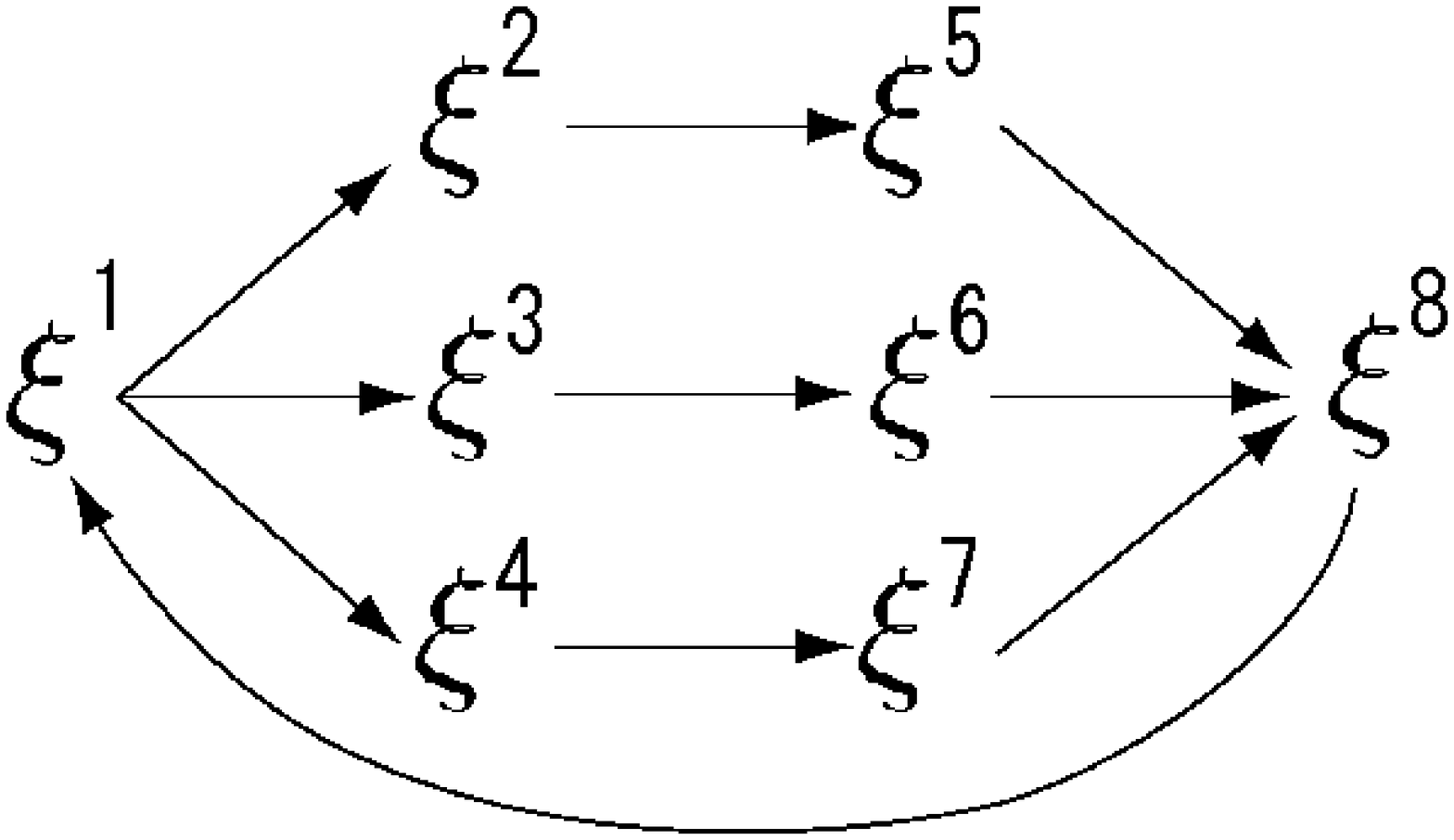} \\ (b) Sequence B. 
\end{center} 
\end{minipage} 
\caption{Two memory sequence patterns. $\xi$ represents memory patterns.} 
\label{fig:seq} 
\end{figure}

The element of the memory patterns $\vec{\xi}^{\mu}=(\xi^{\mu}_1,\cdots,\xi^{\mu}_N)^T$ takes $\pm 1$ with 
\begin{equation} 
\Prob\left[\xi^{\mu}_i=\pm1\right]=\frac{1}{2}. 
\end{equation} 
In this paper, we consider two sequences of memory patterns: sequence B, which is a typical branching sequence, and sequence A, which is a sub-sequence of sequence B (Fig.~\ref{fig:seq}). Synaptic connection $J_{ij}$ is given by 
\begin{equation} 
J_{ij}=\frac1N \sum_{\mu=1}^{p}\sum_{\nu=1}^{p} \xi^{\mu}_i A_{\mu\nu} \xi^{\nu}_j - \frac{p}{N} \, \delta_{ij}. 
\label{eqn:Jij} 
\end{equation} 
The number of patterns following pattern $\vec{\xi}^{\mu}$ is denoted as $p_\mu$. The ($\mu,\nu$)-th component of a matrix $A$ is determined according to 
\begin{align} 
A_{\mu\nu} = \left\{ 
\begin{array}{rl} 1, & \mbox{if $\mu = \nu$} \\ \varepsilon / p_\nu, & \mbox{if $\mu \ne \nu, \ 	\xi^\nu \to\xi^\mu $} \\ 0, & \mbox{if $\mu \ne \nu,\ \xi^\nu \not\to \xi^\mu$} 
\end{array}, \right. 
\end{align} 
Matrix $A$ becomes a transition matrix. For example, in the sequence shown in Fig.~\ref{fig:seq}(a), the numbers of the patterns become $p_1 = 3$, and $p_2 = p_3 = p_4 = 0$, and the matrix becomes 
\begin{equation} 
A = \left( 
\begin{array}{cccc} 1 & 0 & 0 & 0 \\ \varepsilon/3 & 1 & 0 & 0 \\ \varepsilon/3 & 0 & 1 & 0 \\ \varepsilon/3 & 0 & 0 & 1 
\end{array} \right), 
\label{eq:Asammple} 
\end{equation} 
where a parameter $\varepsilon$ is $\varepsilon\ll1$, which determines 
the strength of the cross-correlation connection. With $A_{\mu\nu} = 
\delta_{\mu\nu}$, the model corresponds to an autoassociative memory 
model, whereas with $A_{\mu\nu} = \delta_{\mu,\nu+1}$, it corresponds to 
a sequential associative memory model. 
Kawamura and Okada's model is represented as 
$A_{\mu\nu} = \delta_{\mu\nu} + \varepsilon \delta_{\mu,\nu+1}$\cite{Kawamura2006}.

We define the overlap as the direction cosine between the state of neurons, $\vec{x}^t$ at time $t$, and the memory pattern, $\vec{\xi}^{\mu}$, 
\begin{equation} 
m^{\mu}_t = \frac{1}{N}\sum_{i=1}^N \xi_i^{\mu} x_i^t. 
\label{eqn:mt} 
\end{equation} 
From eq.~(\ref{eqn:dynamics}), (\ref{eqn:Jij}), and (\ref{eqn:mt}), the state of the neurons, $x_i^{t+1}$ becomes 
\begin{align} 
\! x_i^{t+1} \! = \! F \!\left( \sum_{\mu=1}^{p}\sum_{\nu=1}^{p} A_{\mu\nu} \xi^{\mu}_i m^{\nu}_t - \! \frac{p}{N} \, \delta_{ij} + \! \zeta_i^t + \eta^t \! + cB^t_i \! \right). 
\label{eqn:x_xi} 
\end{align} 
The initial state $\vec{x}^0$ is determined according to 
\begin{equation} 
\Prob[x_i^0=\pm1] = \frac{1\pm m_0 \xi^{1}_i}{2}. 
\end{equation} 
Therefore, the overlap between pattern $\vec{\xi}^1$ and initial state $\vec{x}^0$ is $m_0$.

\subsection{Effect of bias input} 
\begin{figure}[tb] 
\begin{center} \includegraphics[width=0.4\linewidth]{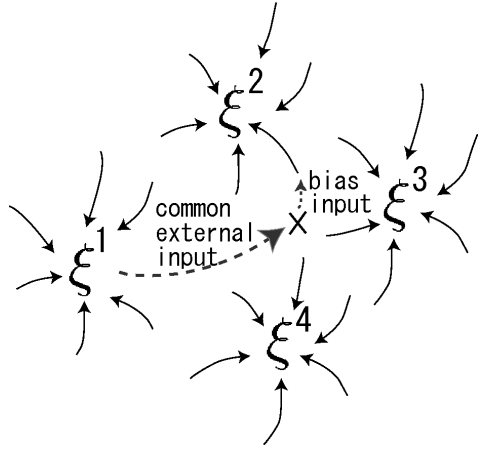} 
\end{center} 
\caption{Schematic diagram of branching state transition mechanisms induced by common external input and bias input. x-marks denote mixed state of patterns $\vec{\xi^{2}}$, $\vec{\xi^{3}}$, and $\vec{\xi^{4}}$.} 
\label{fig:schema} 
\end{figure}

Here, we explain the role of bias input $B^t_i$. We define $h_i^t$ in internal potential as 
\begin{align} 
h_i^t \equiv \sum_{\mu=1}^{p}\sum_{\nu=1}^{p} A_{\mu\nu} \xi_i^{\mu} m^{\nu}_t. 
\end{align} 
When sequence A is stored, from eq.~(\ref{eqn:Jij}) and eq.~(\ref{eq:Asammple}), 
\begin{align} 
h_i^t = \sum_{\mu=1}^p \xi^{\mu}_i m^{\mu} + \frac{\varepsilon}{3} \left(\xi^{2}_i + \xi^{3}_i + \xi^{4}_i\right) m_t^1 . 
\end{align} 
If $\vec{x}^t \approx \vec{\xi}^1$, that is, $m_t^1 \approx 1$, then $h_i^t$ becomes 
\begin{align} 
h_i^t \approx \xi^{1}_i + \frac{\varepsilon}{3} \left(\xi^{2}_i + \xi^{3}_i + \xi^{4}_i\right). 
\end{align} 
The common external input, therefore, will make the state transit to the mixture state $\sgn(\vec{\xi}^{2} + \vec{\xi}^{3} + \vec{\xi}^{4})$ \cite{Kawamura2006}. Since the mixture state correlates with memory patterns $\vec{\xi}^{2}$, $\vec{\xi}^{3}$, and $\vec{\xi}^{4}$, the model has the ability to transit to those three patterns. However, with $N \to \infty$, the distances between the mixed state and the three memory patterns are equal. This symmetry prevents the model from transiting to any memory pattern. In this state, the bias input that correlates with a given memory pattern can break the symmetry and can induce a transition to the target memory pattern (Fig.~\ref{fig:schema}). The value of $b_t^{\nu}$ for the target pattern could be any appropriate small value capable of breaking the symmetry. 

\section{No bias input model ($c = 0$)} 
\subsection{Theory} 
First, we discuss the model without bias input, i.e., $c = 0$. 
In this case, our model can be regarded as a generalized model of Kawamura and 
Okada's one with $A_{\mu\nu} = \delta_{\mu\nu} + \varepsilon \delta_{\mu,\nu+1}$\cite{Kawamura2006}. 
To describe the behavior of the model using macroscopic variables, we consider the thermodynamic limit $N \to \infty$. For the sake of simplicity, we assume that the number of memory patterns is finite, $p {\cal \sim O}(1)$. Independent external input $\zeta_i^t$ is time independent of each neuron and is assumed to obey the Gaussian distribution, where the mean equals $0$ and the variance equals $\sigma^2$. The common external input $\eta^t$ is also time independent, and is assumed to obey the Gaussian distribution, where the mean equals 0 and the variance equals $\delta^2$. 

Due to the common external input, the states of the neurons are correlated, so sample dependence must be taken into account\cite{Amari2003, YamanaOkada2005, Kawamura2005, Kawamura2006}. We derive a probability density function (PDF) of macroscopic variables $\vec{m}_{t}$. When $\eta^t$ is known at given time $t$ and $N \to \infty$, $m^{\mu}_{t+1}, \mu=1,2,...,p$ could be given as the function of $\vec{m}_t = (m^{1}_t, m^{2}_t,..,m^{p}_t)$ and $\eta^t$, 
\begin{align} 
m^{\mu}_{t+1}(\vec{m}_t,\eta^t) &= \frac{1}{N}\sum_{i=1}^N \xi^{\mu}_i F\left(\sum_{\nu=1}^{p}\sum_{\rho=1}^{p} A_{\nu\rho} \xi^{\nu}_i m^{\rho}_t +\zeta^t_i + \eta^t\right) \\ &= \left< \int \!\! D_z \xi^{\mu} F \left(\sum_{\nu=1}^{p}\sum_{\rho=1}^{p} A_{\nu\rho} \xi^{\nu} m^{\rho}_t +\sigma z + \eta^t \right) \right>_{\xi} \\ &= \left< \xi^{\mu} \erf\left( \frac{\sum_{\nu=1}^{p}\sum_{\rho=1}^{p} A_{\nu\rho} \xi^{\nu} m^{\rho}_t	 + \eta^t} {\sqrt{2}\sigma}\right)\right>_{\xi} , 
\label{eqn:merf_c} 
\end{align} 
where $D_{z}=\frac{dz}{\sqrt{2\pi}} \exp\left(-\frac{z^2}{2}\right)$ and $\left<\cdot\right>_{\xi}$ denotes the average over the whole memory pattern $\vec{\xi}^{\mu}$.  The function $\erf(u)$ is defined as 
\begin{equation} 
\erf\left(u\right)=\frac{2}{\sqrt{\pi}}\int_0^udt \exp\left(-t^2\right). 
\end{equation} 
Next, when $\eta^t$ obeys the Gaussian distribution: 
\begin{equation} 
p\left(\eta^t\right) = \frac{1}{\sqrt{2\pi}\delta} \exp\left(-\frac{\left(\eta^t\right)^2}{2\delta^2}\right), 
\end{equation} 
the distribution of overlap $\vec{m}_t$ can be described as the PDF, $p\left(\vec{m}_t,\eta^t\right)$. Since $\eta^t$ is independent of $\vec{m}_t$, $p\left(\vec{m}_t,\eta^t\right)$ can be divided into two PDFs, 
\begin{equation} 
p\left(\vec{m}_t,\eta^t\right) = p\left(\vec{m}_t\right)p\left(\eta^t\right). 
\label{eqn:independ} 
\end{equation} 
Therefore, the PDF of $\vec{m}_t$ is given by 
\begin{align} 
p\left(\vec{m}_{t+1}\right) &= \int \prod_{\nu=1}^p dm^{\nu}_t d\eta^t p\left(\vec{m}_t\right) p\left(\eta^t\right) \prod_{\nu=1}^p \delta\left(m^{\nu}_{t+1}-m^{\nu}_{t+1}(\vec{m}_t,\eta^t)\right), 
\label{eqn:Pnext} 
\end{align} 
where $\delta(\cdot)$ denotes the Dirac delta function. We evaluate (\ref{eqn:Pnext}) using the Monte Carlo method in the following section.

\subsection{Results} 
\begin{figure}[tb] 
\begin{center} \includegraphics[width=0.4\linewidth]{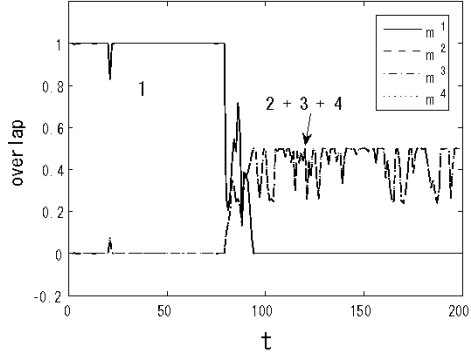} 
\end{center} 
\caption{Time evolutions of overlaps $m^{1}_t,m^{2}_t, m^{3}_t$, and $m^{4}_t$ obtained theoretically for bias input ($c=0$), where $\varepsilon=0.1, \sigma=0.1$, and $\delta=0.37$. } 
\label{fig:ovlp_theory_nobias} 
\end{figure}

Focusing on branching state transitions, we mainly treat sequence A in 
Fig.~\ref{fig:seq}(a). Figure \ref{fig:ovlp_theory_nobias} shows a 
sample of time evolution of the overlap by (\ref{eqn:merf_c}), where 
$\varepsilon = 0.1$, $\sigma=0.1$, and $\delta=0.37$. The initial overlap is $m_0 = 1.0$. Numbers in Fig.~\ref{fig:ovlp_theory_nobias} denote indexes of the memory patterns. Because there is no bias input, $m^{2}_t, m^{3}_t$, and $m^{4}_t $ behave in a similar fashion. Therefore, no memory pattern is retrieved completely; only a mixed state is retrieved.

\begin{figure}[tb] 
\begin{center} \includegraphics[width=0.50\linewidth]{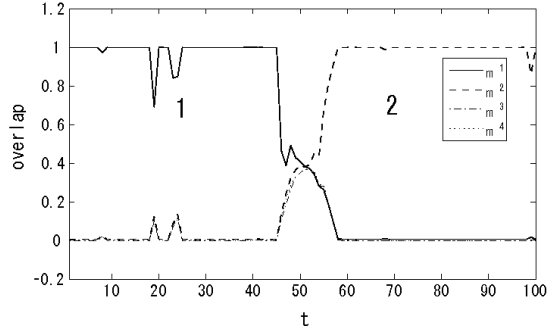} 
\end{center} 
\caption{Time evolution of overlaps $m^{1}_t,m^{2}_t, m^{3}_t$, and $m^{4}_t $ when there is no bias input ($c=0$), obtained by computer simulations ($N=100,000$), where $\varepsilon=0.1, \sigma=0.1$, and $\delta=0.37$. State transition occurs from pattern $\vec{\xi}^1$ to $\vec{\xi}^2$} 
\label{fig:ovlp_no_biasl_sim} 
\end{figure}

Figure~\ref{fig:ovlp_no_biasl_sim} shows a sample of the time evolution 
of overlaps calculated by computer simulation, where $\varepsilon = 0.1$, 
$\sigma=0.1$, $\delta=0.37$, and the number of neurons is $N=100,000$. The transition to the mixed state occurs after the pattern $\vec{\xi}^1$. In this sample, even though there is no bias input, pattern $\vec{\xi}^4$ is completely retrieved. The memory patterns $\vec{\xi}^3$ and $\vec{\xi}^4$ can also be retrieved in other samples. 

\begin{figure}[tb] 
\begin{minipage}{.50\linewidth} 
\begin{center} \includegraphics[width=75mm]{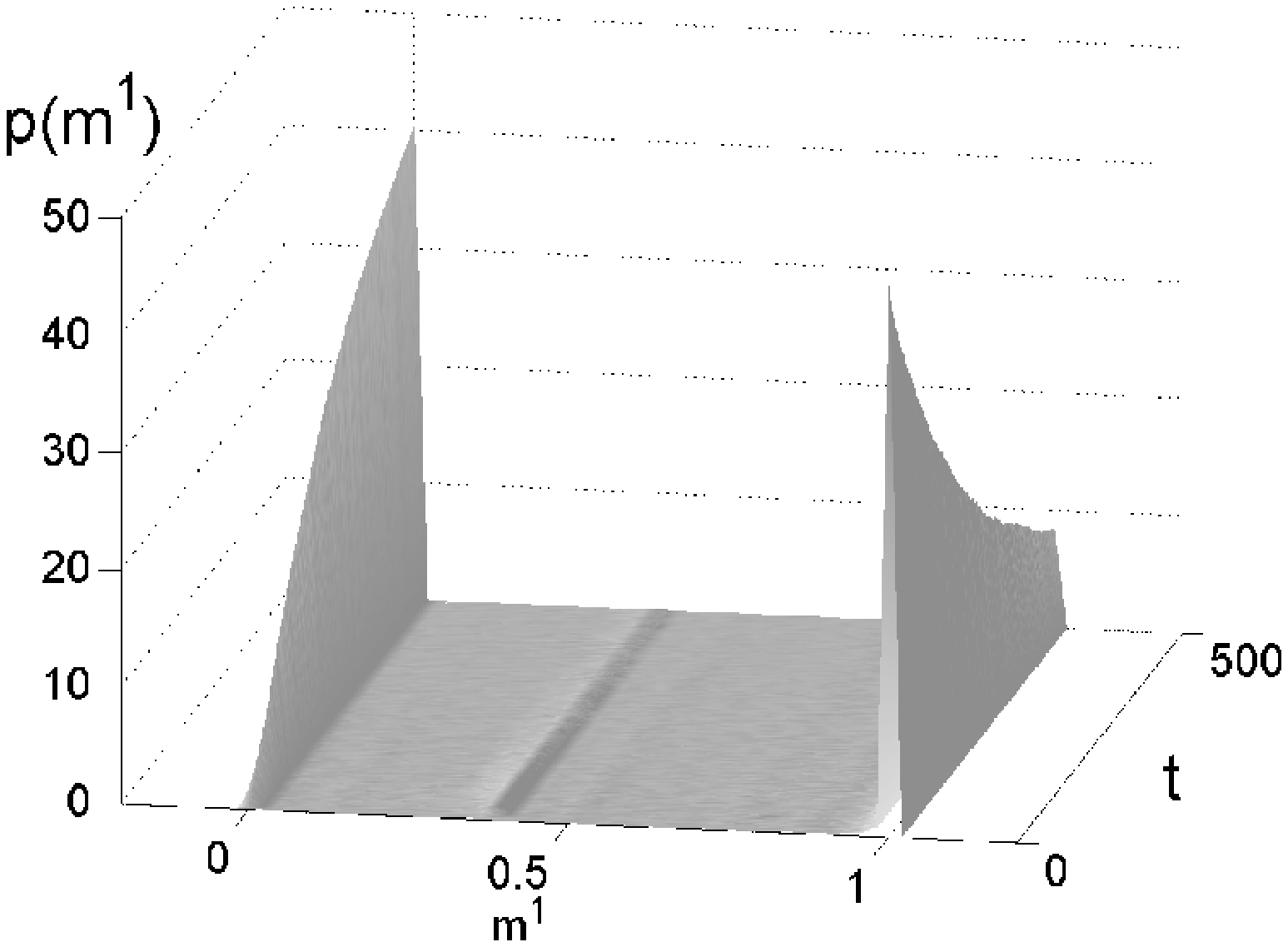} \\ (a) $m^1$ 
\end{center} 
\end{minipage} 
\begin{minipage}{.50\linewidth} 
\begin{center} \includegraphics[width=75mm]{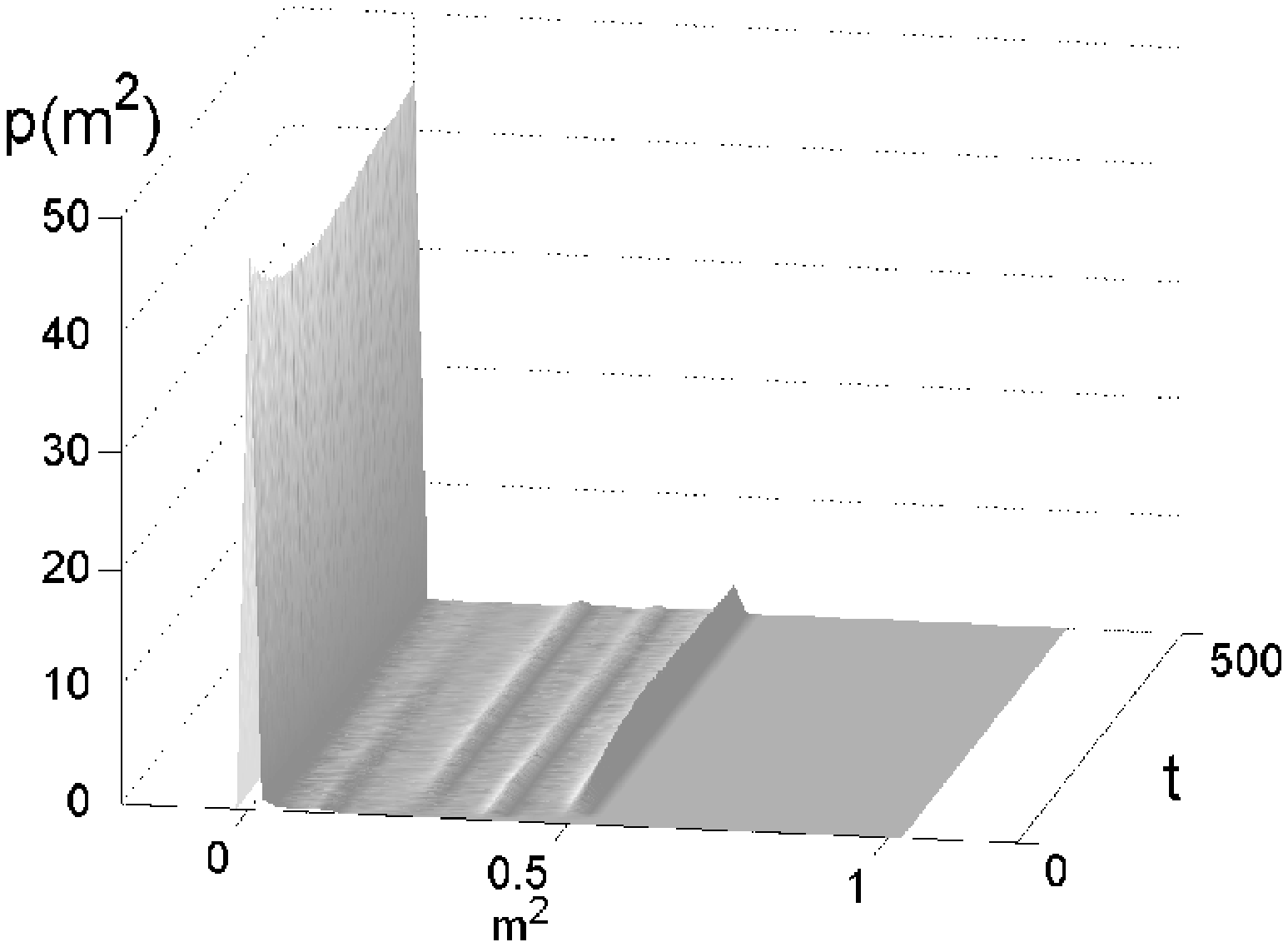} \\ (b) $m^2$ 
\end{center} 
\end{minipage} \\ \\ 
\begin{minipage}{.50\linewidth} 
\begin{center} \includegraphics[width=75mm]{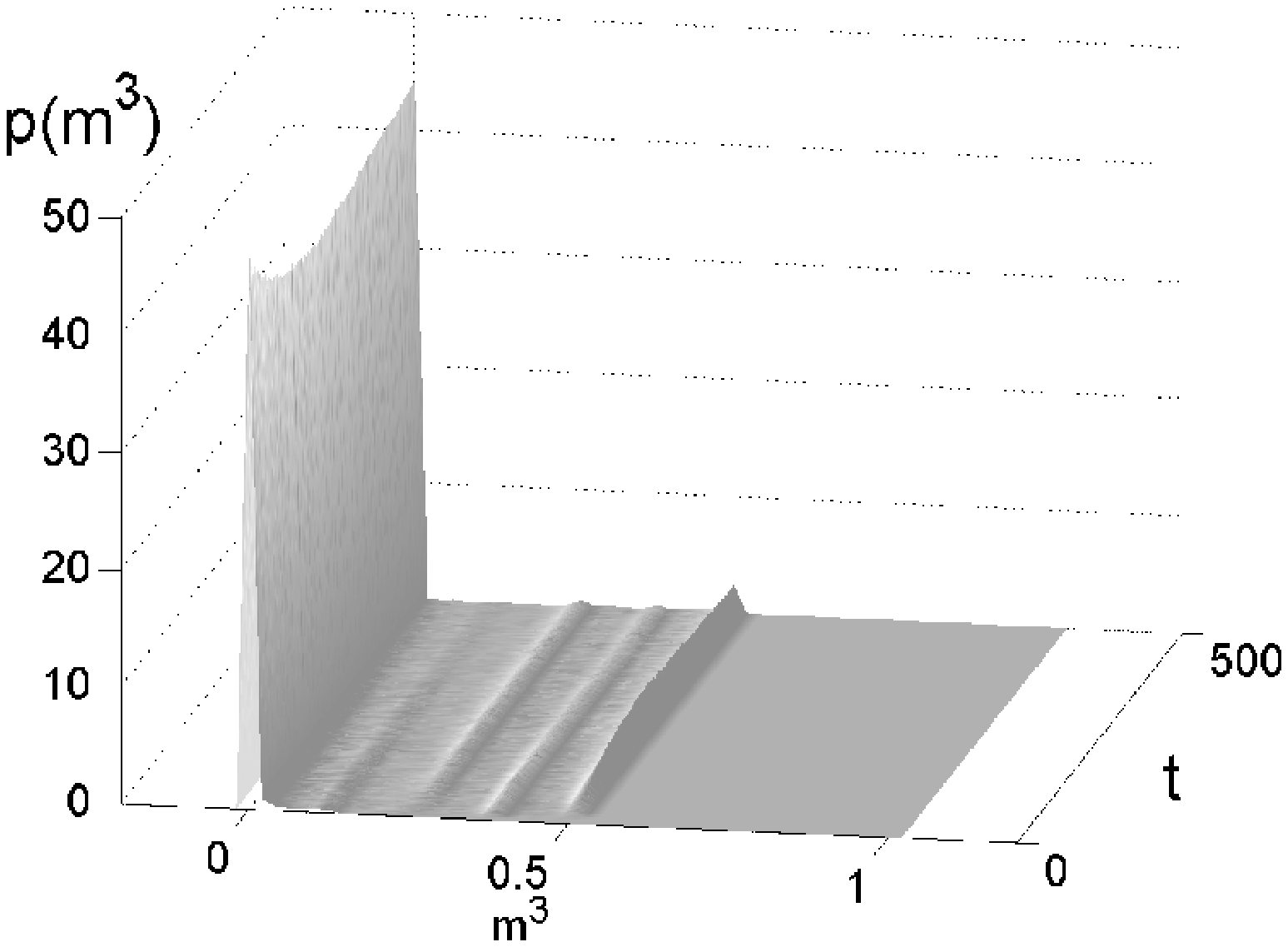} \\ (c) $m^3$ 
\end{center} 
\end{minipage} 
\begin{minipage}{.50\linewidth} 
\begin{center} \includegraphics[width=75mm]{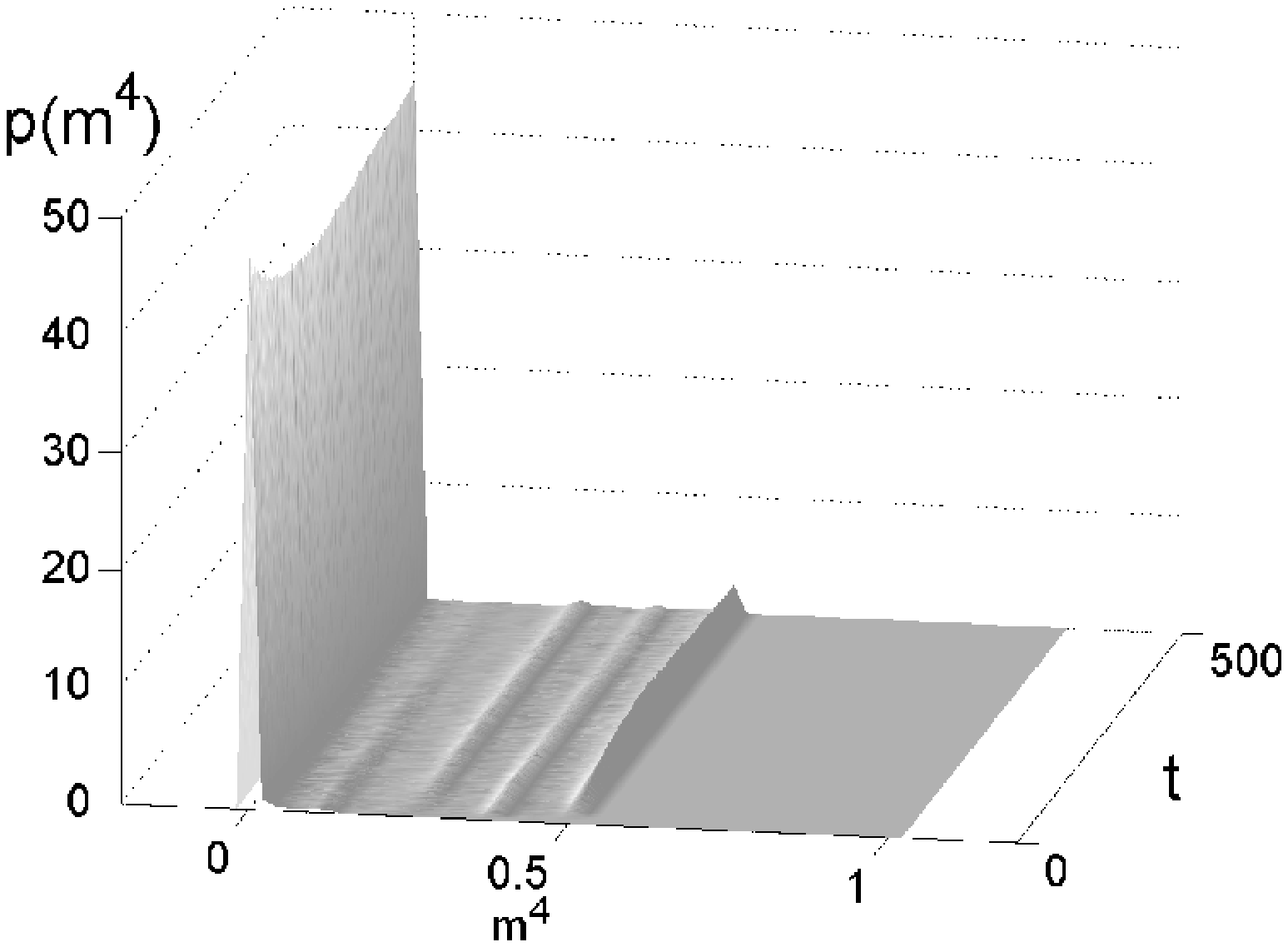} \\ (d) $m^4$ 
\end{center} 
\end{minipage} 
\caption{ PDFs of overlaps obtained theoretically in model without bias input ($c = 0$), where $\varepsilon=0.1, \sigma=0.1$, and $\delta=0.37$.} 
\label{fig:pdf_nobias} 
\end{figure}

\begin{figure}[tb] 
\begin{minipage}{.50\linewidth} 
\begin{center} \includegraphics[width=75mm]{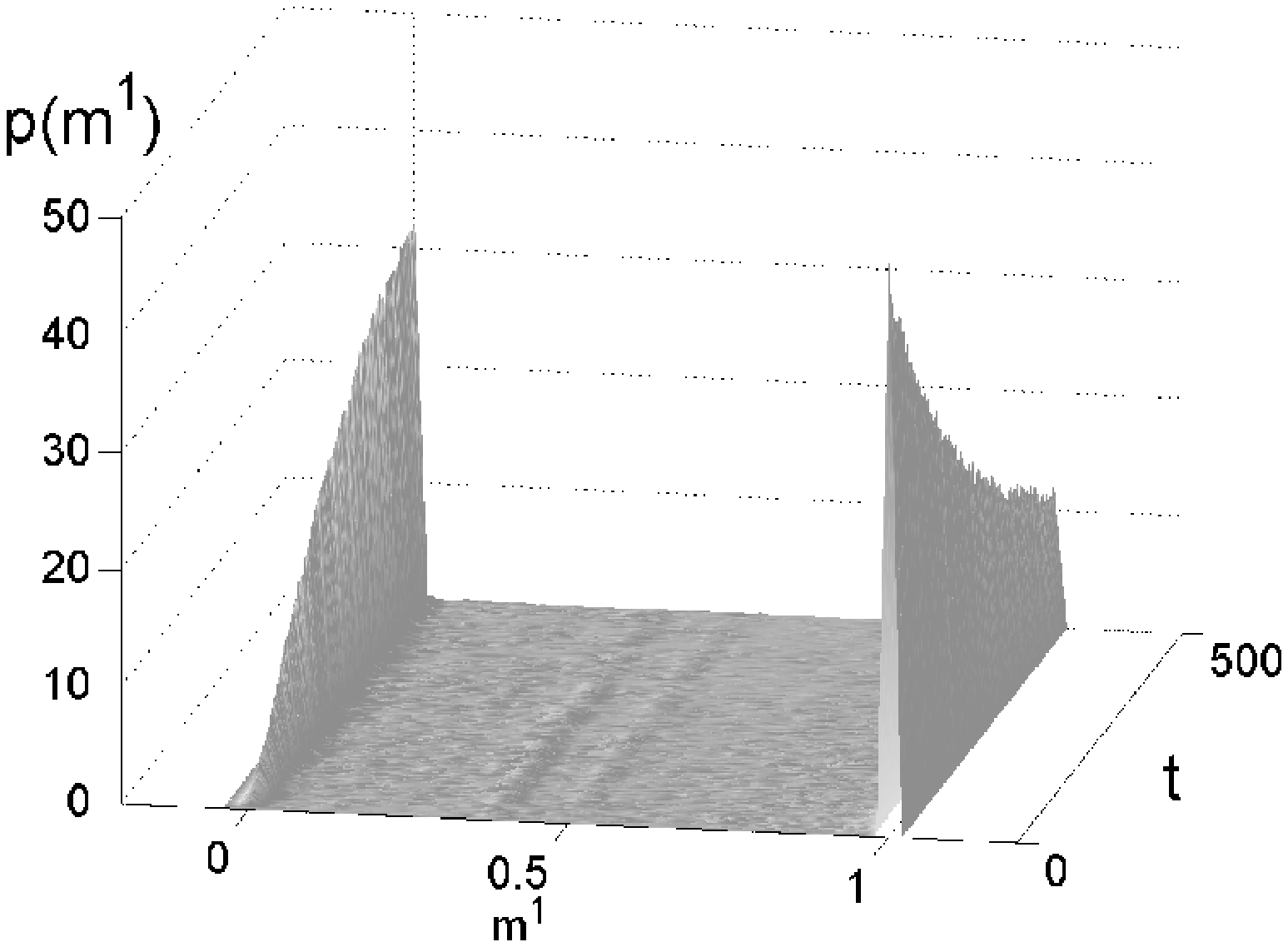} \\ (a) $m^1$ 
\end{center} 
\end{minipage} 
\begin{minipage}{.50\linewidth} 
\begin{center} \includegraphics[width=75mm]{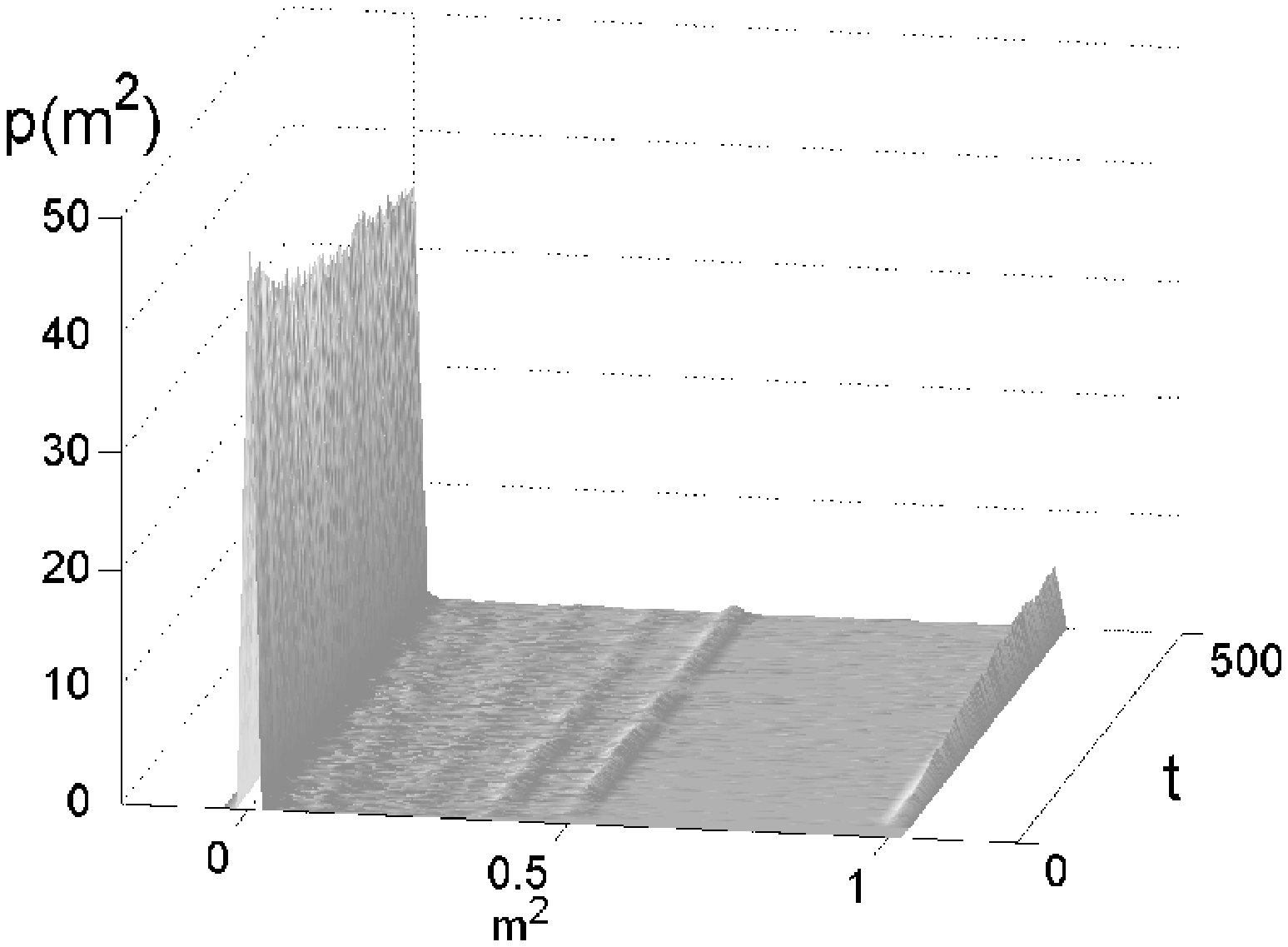} \\ (b) $m^2$ 
\end{center} 
\end{minipage} \\ 
\begin{minipage}{.50\linewidth} 
\begin{center} \includegraphics[width=75mm]{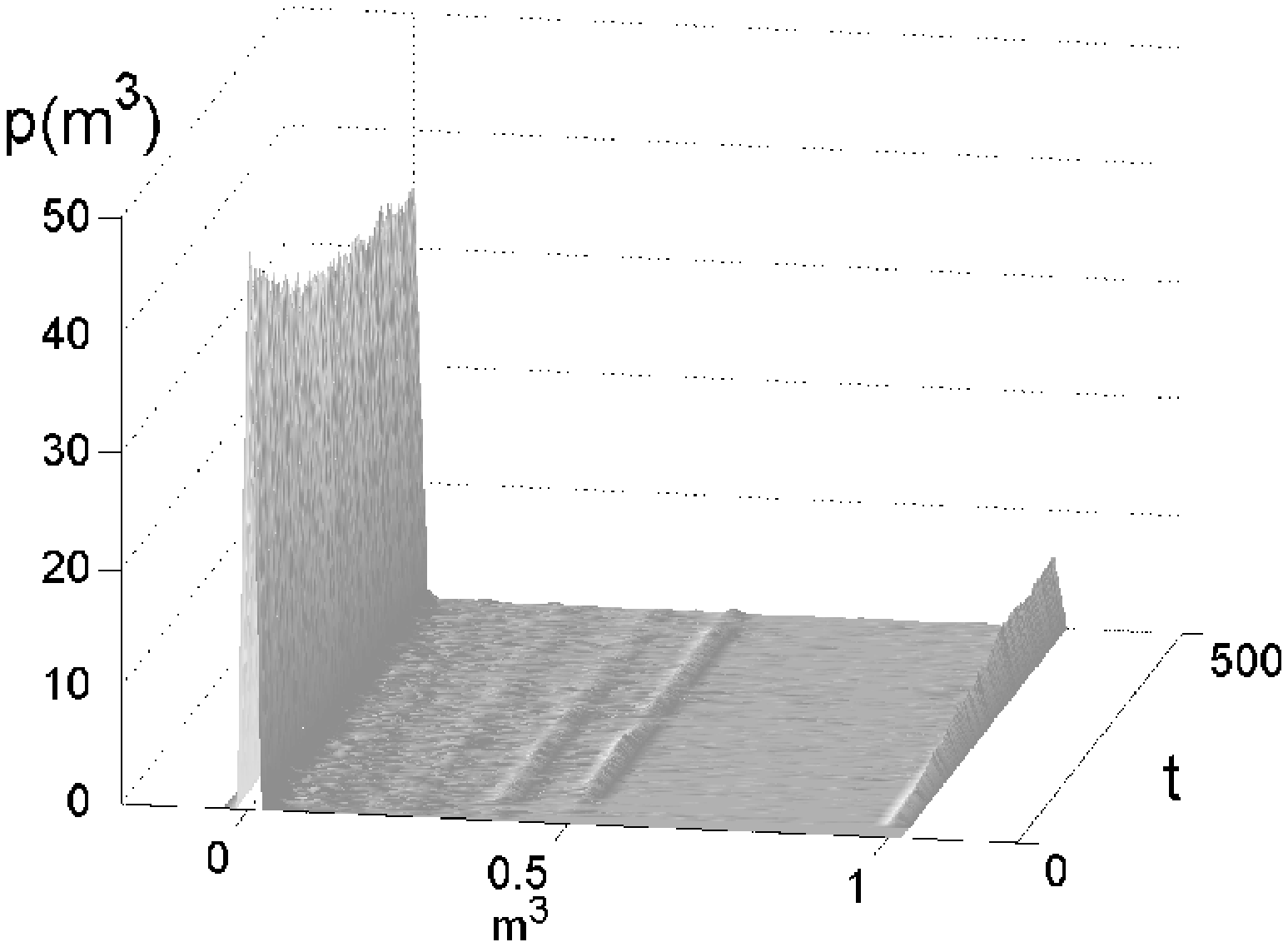} \\ (c) $m^3$ 
\end{center} 
\end{minipage} 
\begin{minipage}{.50\linewidth} 
\begin{center} \includegraphics[width=75mm]{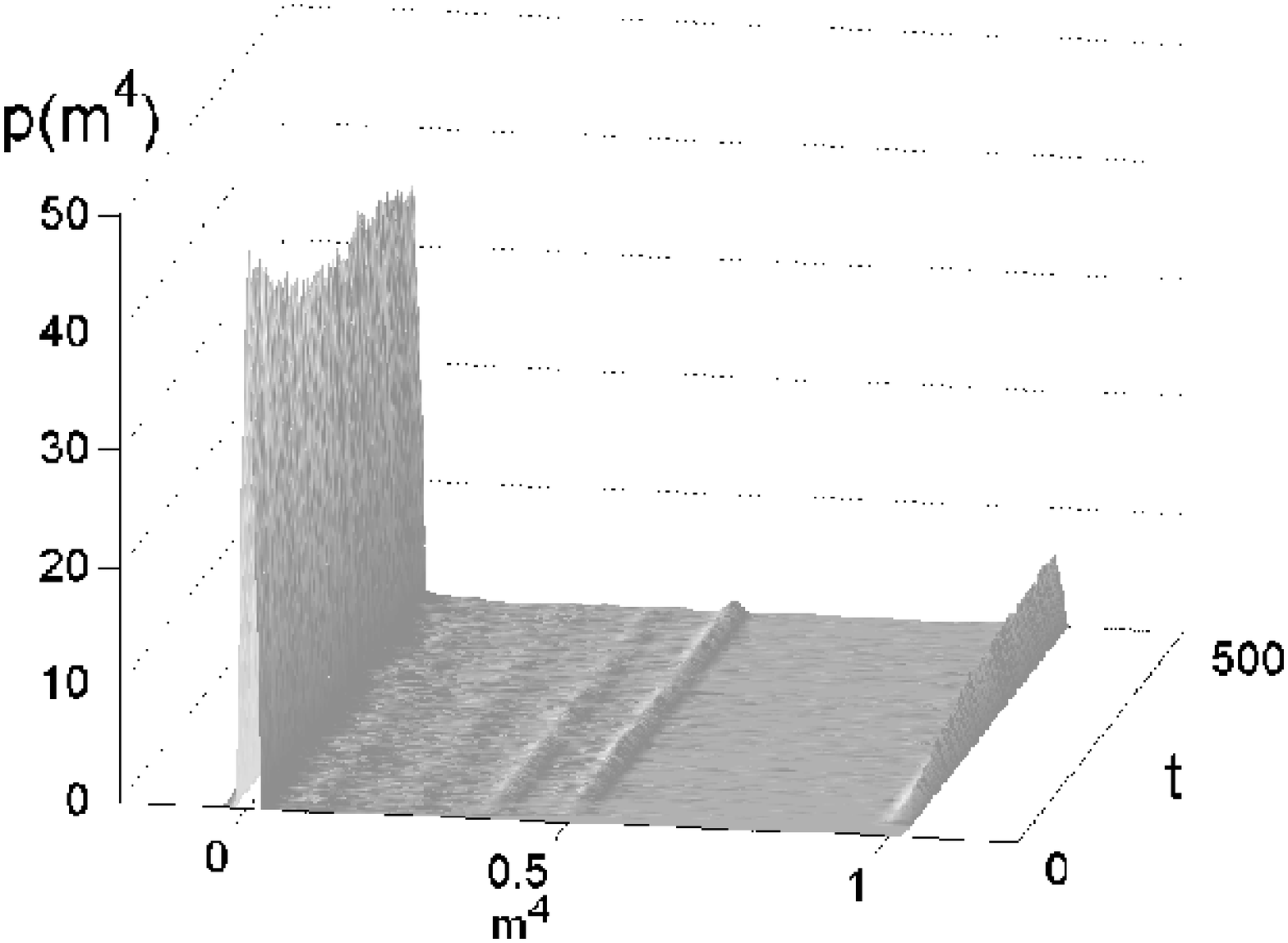} \\ (d) $m^4$ 
\end{center} 
\end{minipage} 
\caption{Histograms of overlaps obtained by computer simulations ($N=100,000$). Parameters are same as in Fig.\ref{fig:pdf_nobias}.} 
\label{fig:hist_nobias} 
\end{figure}

Since the states of neurons have sample dependence due to common external input, we investigate the distributions of the overlaps. To evaluate the probability of a given memory pattern being retrieved, we calculate the marginal probability density functions of $m^{\mu}_t$, $\mu=1,2,3$, and $4$, from the joint probability density function $p\left(\vec{m}_t\right)$ in eq.~(\ref{eqn:Pnext}), 
\begin{equation} 
p(m^{\mu}_{t})=\int \prod_{\nu\neq\mu}dm^{\nu}_t\; p\left(\vec{m}_t\right). 
\label{eqn:Pmt} 
\end{equation} 
Figure~\ref{fig:pdf_nobias} shows the marginal PDFs of the overlaps, $p\left(m^1_t\right)$, $p\left(m^2_t\right)$, $p\left(m^3_t\right)$, and $p\left(m^4_t\right)$. The probability densities at $m^2 = 1.0$, $m^3=1.0$, and $m^4=1.0$ are zero. That is, the patterns $\vec{\xi}^2, \vec{\xi}^3,$ and $\vec{\xi}^4$ are not retrieved. Figure~\ref{fig:hist_nobias} shows histograms obtained by computer simulation. The frequencies at $m^2 = 1.0, m^3=1.0, m^4=1.0$ are observed to a certain extent. The disagreement between computer simulations and theory may be due to the finite size effect, which breaks the symmetry between the mixed state and the memory patterns. 

\section{Bias input model ($c > 0$)} 
\subsection{Theory} 
Next, we discuss the model with bias input, where $m^{\mu}_{t+1}, \mu=1,2,...,p$ are given by the function of $\vec{m}_t = (m^{1}_t, m^{2}_t,..,m^{p}_t)$ and $\eta^t$, 
\begin{align} 
\lefteqn{m^{\mu}_{t+1}(\vec{m}_t,\eta^t)} \nonumber\\ &= \frac{1}{N}\sum_{i=1}^N \xi^{\mu}_i F\left(h^t +\zeta^t_i + \eta^t + c B^t_i\right) \\ &= \! \Biggl< \! \int \!\! D_z \xi^{\mu} \Biggl\{ \frac{1 + \sum_{\nu=1}^p b_t^\nu \xi^{\nu}}{2} F\left[h^t +\sigma z + \eta^t + c \right] + \, \frac{1 - \sum_{\nu=1}^p b_t^\nu \xi^{\nu}}{2} \, F\left[h^t+\sigma z + \eta^t - c\right] \Biggl\} \Biggl>_{\xi} \\ &= \Biggl< \xi^{\mu} \Biggl\{ 	\frac{1+\sum_{\nu=1}^p b_t^\nu \xi^{\nu}}{2} \, 	 \erf \left( \frac{h^t + \eta^t + c}{\sqrt{2}\sigma} \right) 	+ \frac{1-\sum_{\nu=1}^p b_t^\nu \xi^{\nu}}{2} \, \erf \left( \frac{h^t + \eta^t - c}{\sqrt{2}\sigma} \right) \Biggl\} \, \Biggl>_{\xi} , 
\label{eqn:merf_bias} 
\end{align} 
where 
\begin{align} 
h^t \equiv \sum_{\nu=1}^{p}\sum_{\rho=1}^{p} A_{\nu\rho} \xi^{\nu} m^{\rho}_t. 
\end{align} 
When the memory pattern $\vec{\xi}^2$ is biased ($b^t_2 > 0$, $b^t_1 = b^t_3 = b^t_4 = 0$, and $c > 0$) in network storing sequence A, $m^{\mu}_{t+1}(\vec{m}_t,\eta^t)$ for $\mu \ne 2$ become 
\begin{align} 
m^{\mu}_{t+1}(\vec{m}_t,\eta^t) &= \Biggl< \! \frac{\xi^{\mu}}{2} \Biggl\{\erf \left( \! \frac{h^t + \eta^t + c}{\sqrt{2}\sigma} \! \right) + \erf \left( \! \frac{h^t + \eta^t - c}{\sqrt{2}\sigma} \! \right) 			 \! \Biggl\} \! \Biggr>_{\xi}, 
\end{align} 
and for $\mu = 2$, 
\begin{align} 
m^{2}_{t+1}(\vec{m}_t,\eta^t) &= \Biggl< \frac{\xi^{2}}{2} \Biggl\{\erf \left( \frac{h^t + \eta^t + c}{\sqrt{2}\sigma} \right) + \, \erf \left( \frac{h^t + \eta^t - c}{\sqrt{2}\sigma} \right)  \Biggl\} \nonumber\\ 	 & \ \ \ \ \ \ + \! \frac{b_t^2}{2} \Biggl\{\erf \left( \frac{h^t + \eta^t + c}{\sqrt{2}\sigma} \right) - \erf \left( \frac{h^t + \eta^t - c}{\sqrt{2}\sigma} \right) \!  \! \Biggl\} \! \Biggr>_{\xi} . 
\label{eqn:merf_bias_seq_a} 
\end{align} 
The effect of the bias input appears in the second term of RHS in (\ref{eqn:merf_bias_seq_a}). Therefore, $m_{t}^2$ becomes larger than $m_{t}^3$ and $m_{t}^4$. The PDFs of overlaps $m_t^{\mu}$, $\mu=1,2,...,p$ are also given by the form of eq.~(\ref{eqn:Pnext}). 

\subsection{Results} 
\begin{figure}[tb] 
\begin{center} \includegraphics[width=0.4\linewidth]{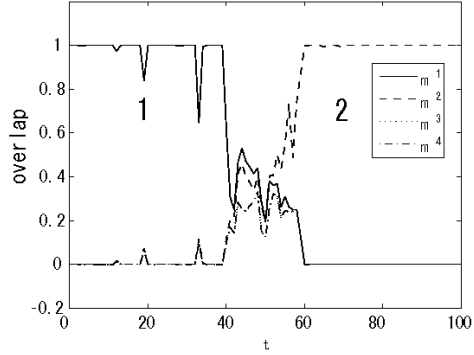} 
\end{center} 
\caption{Time evolution of overlaps obtained theoretically in model with bias input ($b^t_2 = 0.1, c = 0.05$), where $\varepsilon=0.1, \sigma=0.1$, and $\delta=0.37$.} 
\label{fig:ovlp_bias_theory} 
\end{figure}

\begin{figure}[tb] 
\begin{minipage}{.50\linewidth} 
\begin{center} \includegraphics[width=75mm]{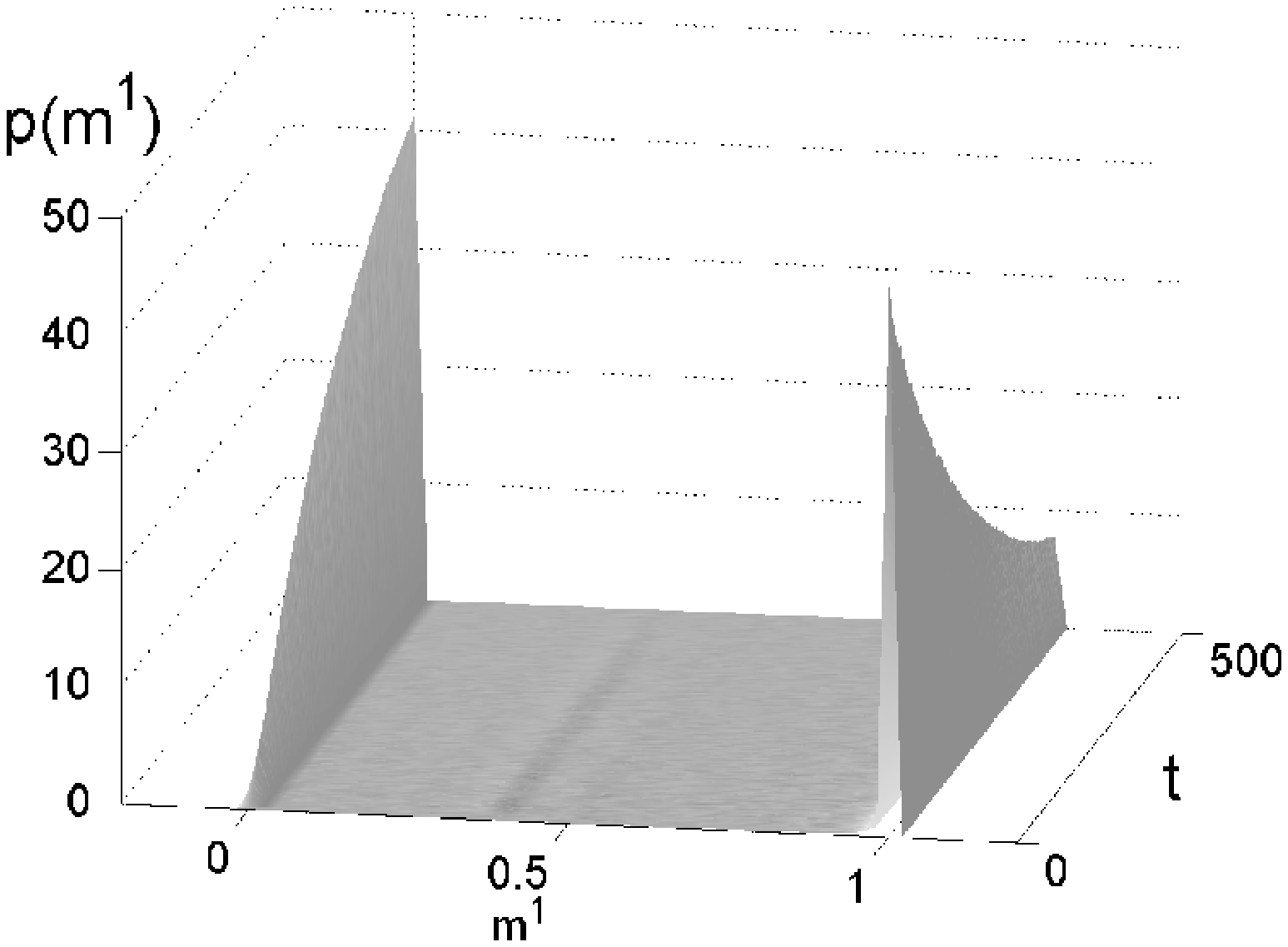} \\ (a) $m^1$ 
\end{center} 
\end{minipage} 
\begin{minipage}{.50\linewidth} 
\begin{center} \includegraphics[width=75mm]{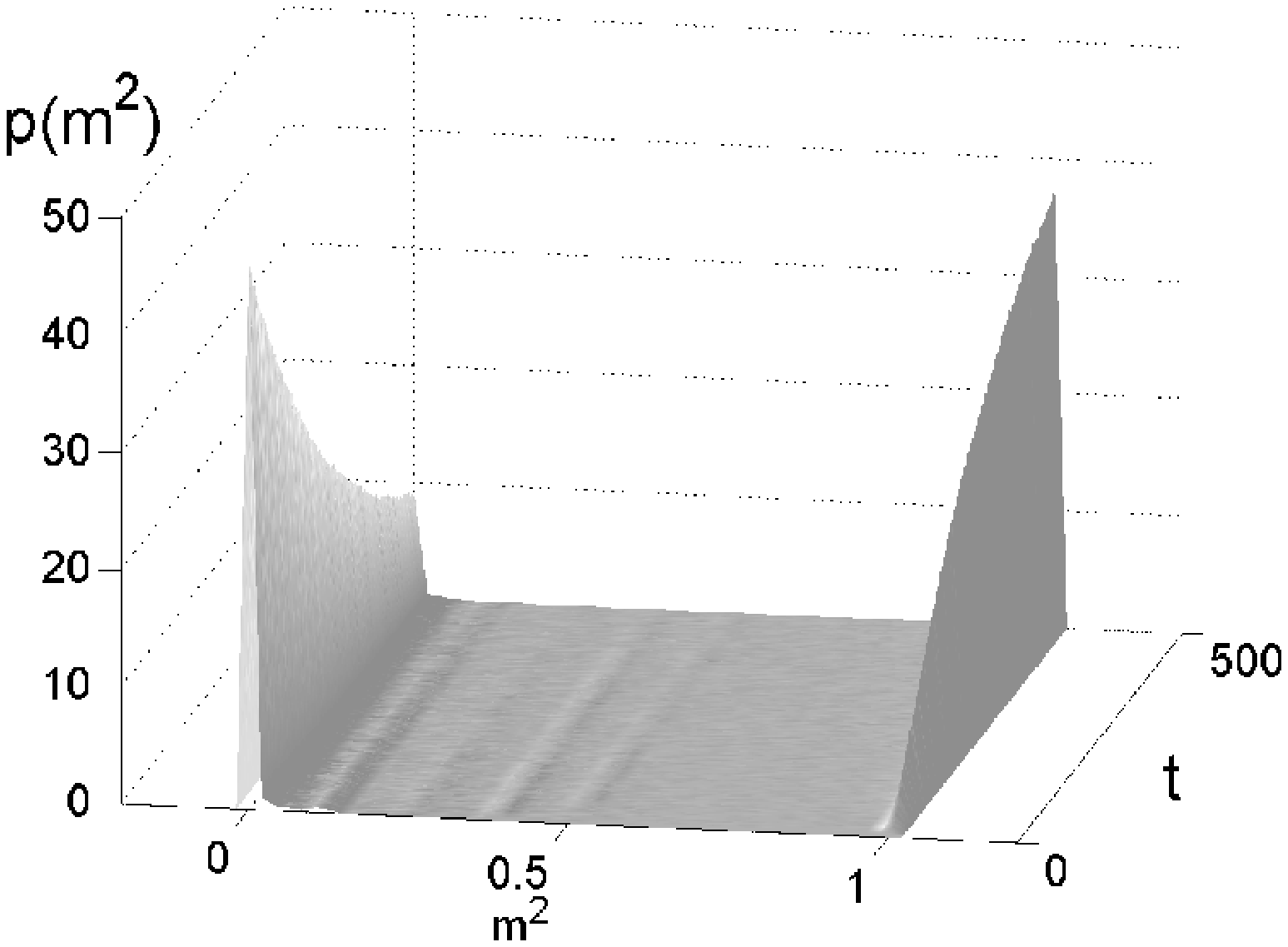} \\ (b) $m^2$ 
\end{center} 
\end{minipage} \\ 
\begin{minipage}{.50\linewidth} 
\begin{center} \includegraphics[width=75mm]{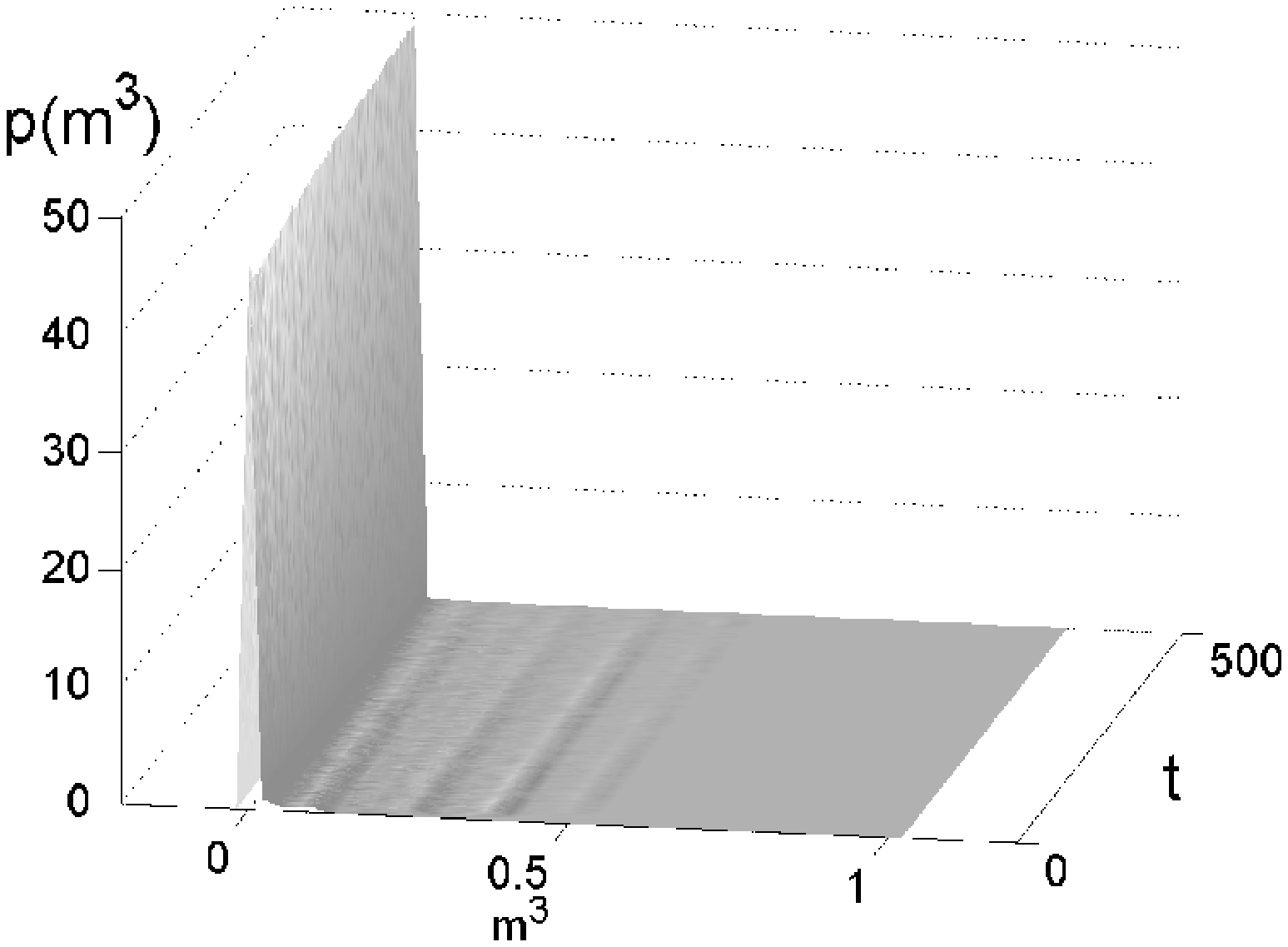} \\ (c) $m^3$ 
\end{center} 
\end{minipage} 
\begin{minipage}{.50\linewidth} 
\begin{center} \includegraphics[width=75mm]{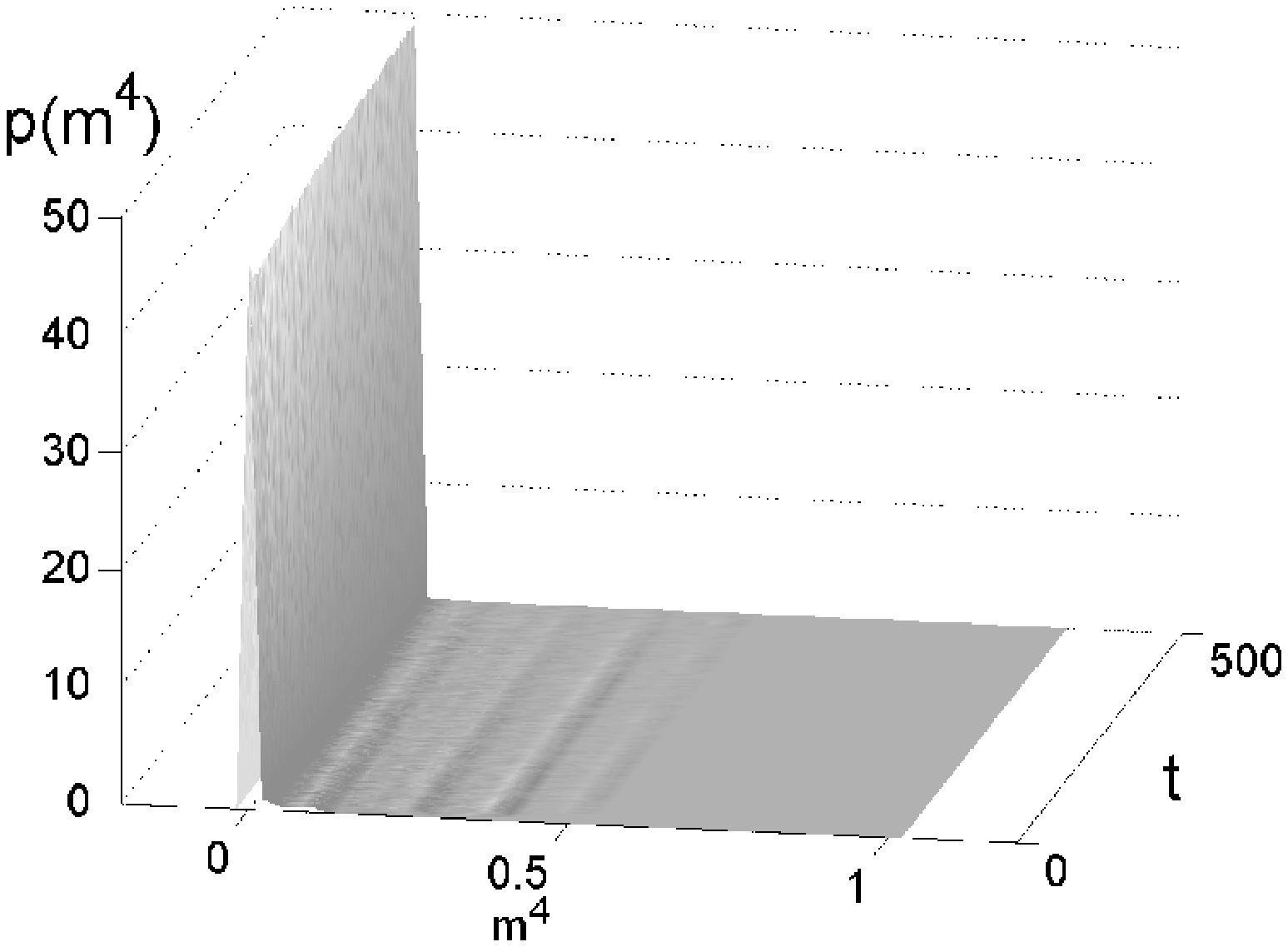} \\ (d) $m^4$ 
\end{center} 
\end{minipage} 
\caption{ PDFs of overlaps obtained theoretically in model with bias input ($b_t^2 = 0.1$, $c = 0.05$), where $\varepsilon=0.1, \sigma=0.1$, and $\delta=0.37$.} 
\label{fig:pdf_noisybias} 
\end{figure}

\begin{figure}[tb] 
\begin{minipage}{.50\linewidth} 
\begin{center} \includegraphics[width=75mm]{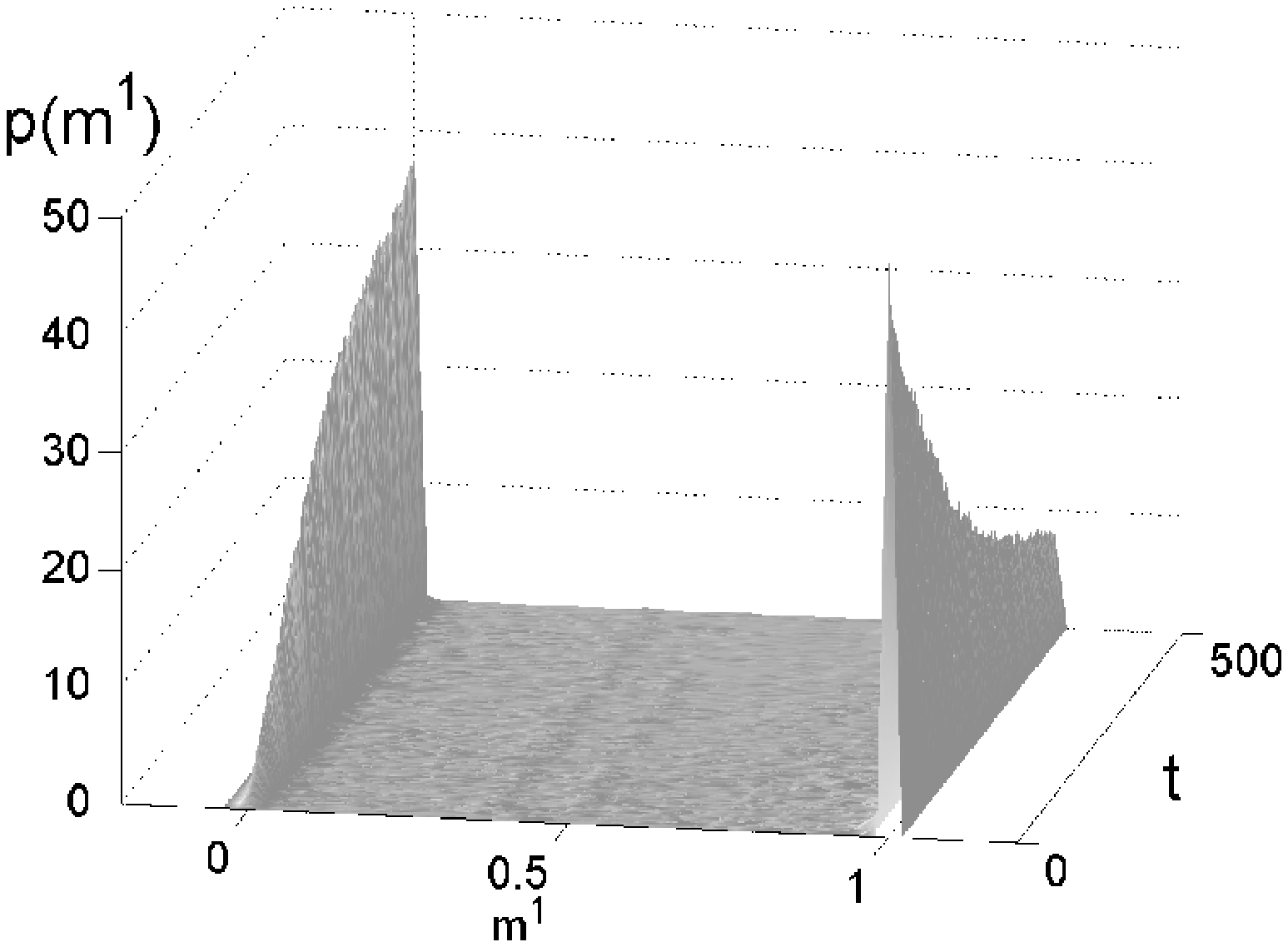} \\ (a)$m^1$ 
\end{center} 
\end{minipage} 
\begin{minipage}{.50\linewidth} 
\begin{center} \includegraphics[width=75mm]{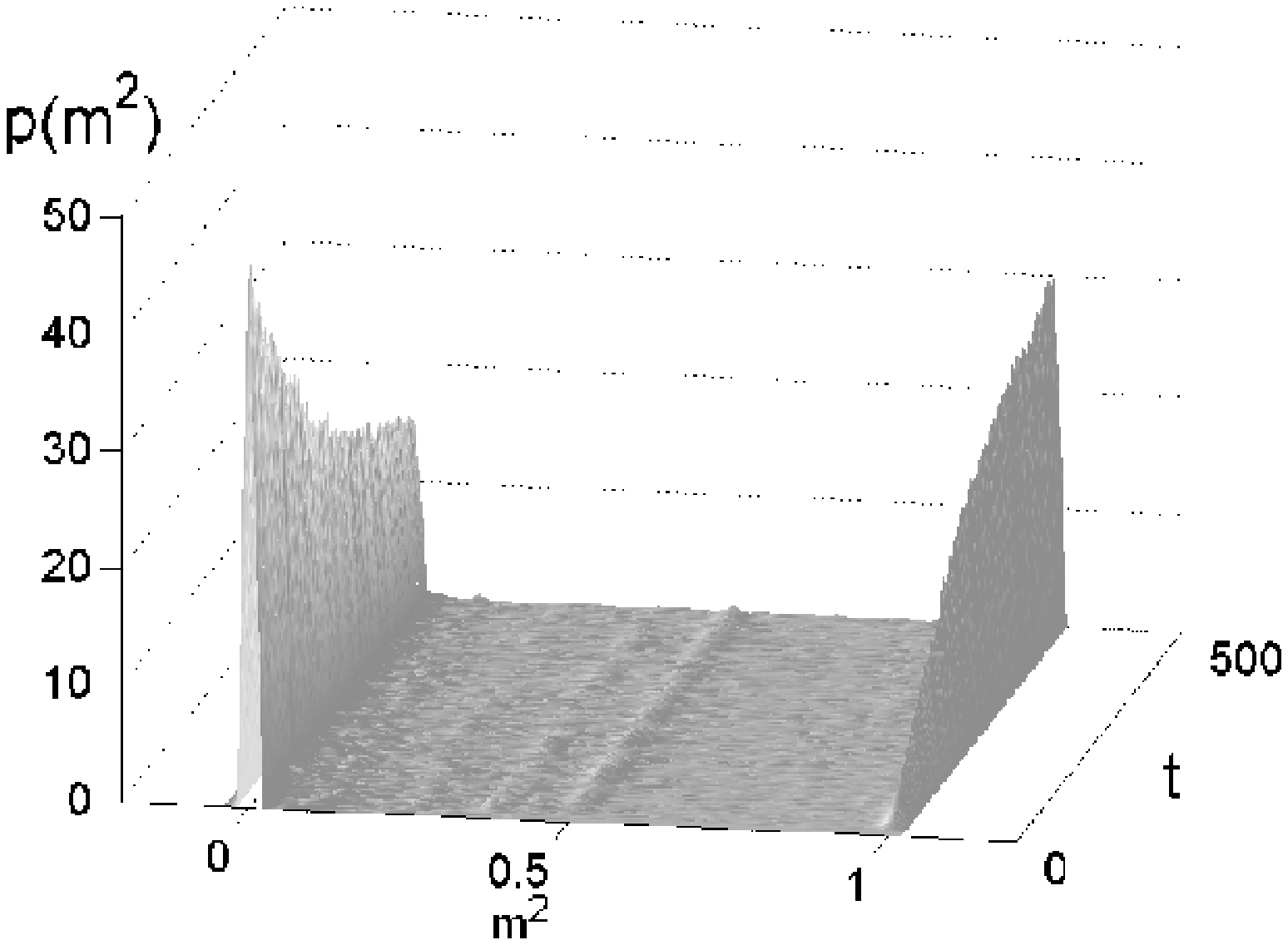} \\ (b)$m^2$ 
\end{center} 
\end{minipage} \\ 
\begin{minipage}{.50\linewidth} 
\begin{center} \includegraphics[width=75mm]{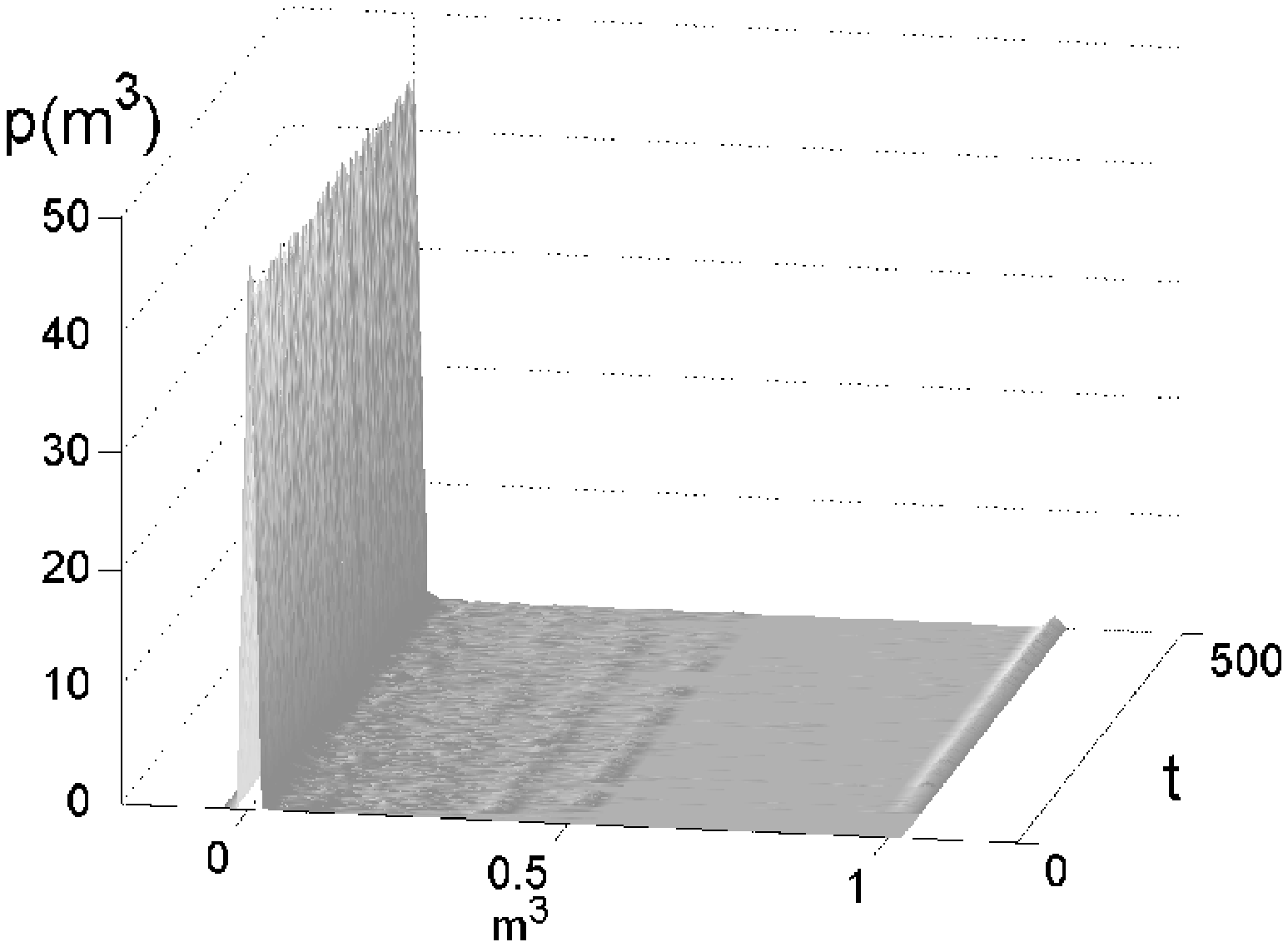} \\ (c)$m^3$ 
\end{center} 
\end{minipage} 
\begin{minipage}{.50\linewidth} 
\begin{center} \includegraphics[width=75mm]{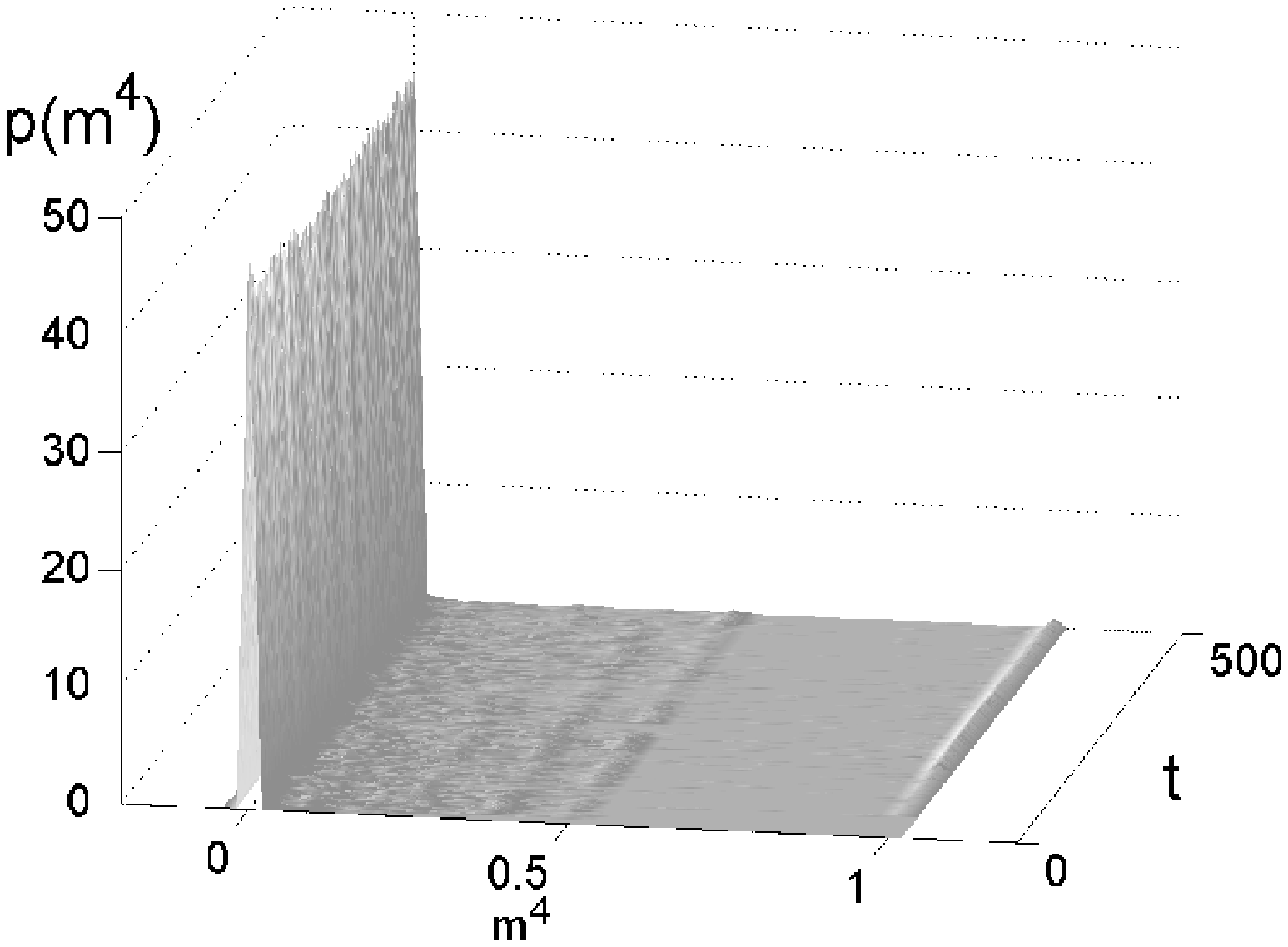} \\ (d)$m^4$ 
\end{center} 
\end{minipage} 
\caption{Histograms of overlaps obtained by computer simulations ($N=100,000$) in model with bias input ($b_t^2 = 0.1$, $c = 0.05$). Parameters are same as in Fig.~\ref{fig:pdf_noisybias}.} 
\label{fig:hist_noisybias} 
\end{figure}

We have shown that the model can transit to one of the memory states, 
i.e., transfer out of a mixed state. The parameters related to the bias 
input are $b_t^2 = 0.1$, $b_t^1 = b_t^3 = b_t^4 = 0$, and $c = 0.05$. 
That is, the bias input, whose overlap with the pattern $\vec{\xi}^2$ is 
$b_t^2 = 0.1$, is multiplied by 0.05. Other parameters are $\varepsilon 
= 0.1$, $\sigma=0.1$, and $\delta=0.37$. Figure~\ref{fig:ovlp_bias_theory} shows the time evolutions of the overlaps $m^{1}_t,m^{2}_t, m^{3}_t$, and $m^{4}_t $ calculated by (\ref{eqn:merf_bias}). Biased pattern $\vec{\xi}^2$ can be retrieved in most samples. Figure~\ref{fig:pdf_noisybias} shows marginalized PDFs $p\left(m^1_t\right)$, $p\left(m^2_t\right)$, $p\left(m^3_t\right)$, and $p\left(m^4_t\right)$ at time $t$. Probability density at $m^2 = 1.0$ increases with time. Figure~\ref{fig:hist_noisybias} shows histograms obtained by computer simulations ($N=100,000$). The model transits to state $m^3 = 1.0$ or $m^4 = 1.0$ with small probability. Except for this small disagreement, the results agree with those obtained theoretically in Fig.~\ref{fig:pdf_noisybias}.

\begin{figure}[tbh] 
\begin{center} \includegraphics[width=0.7\linewidth]{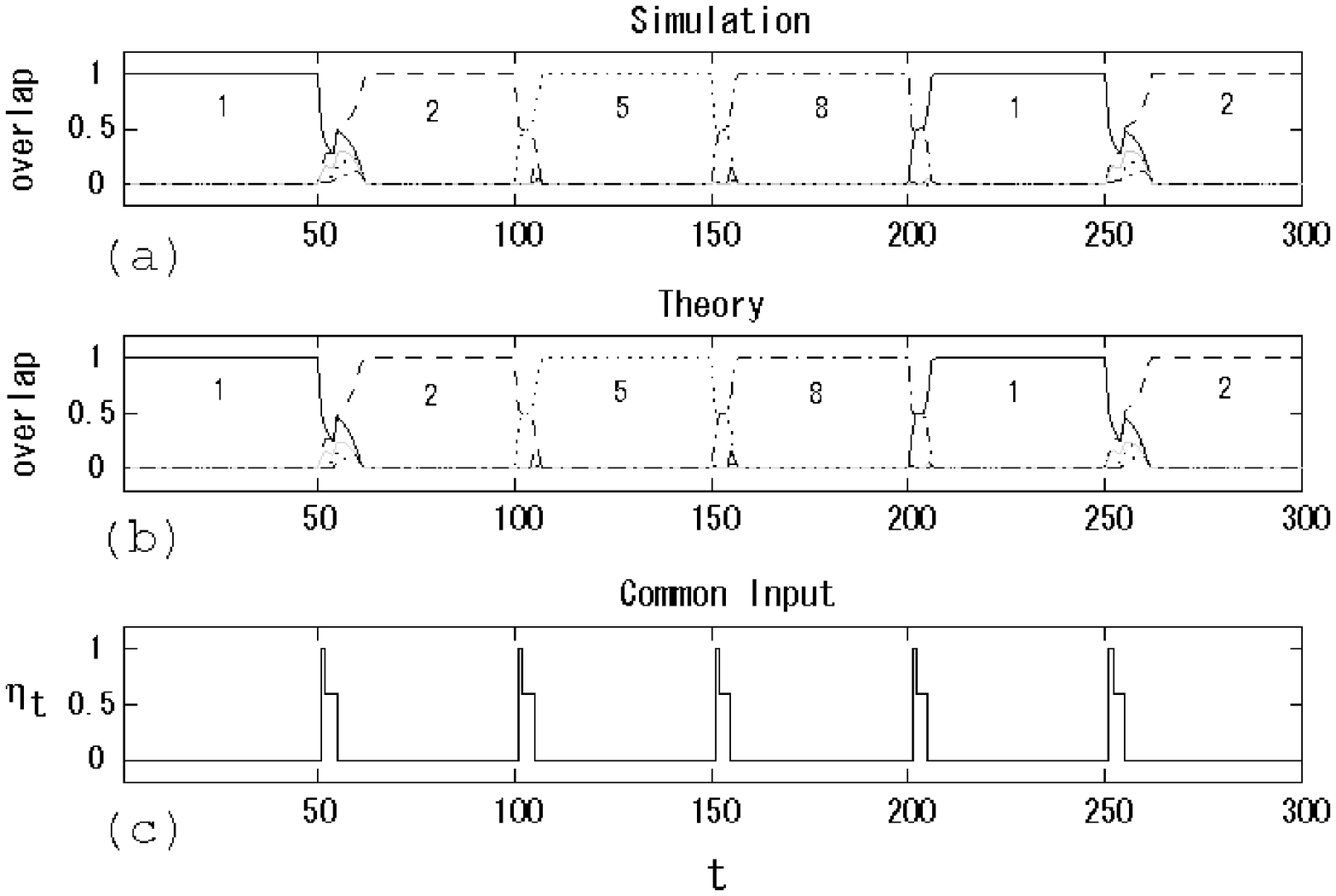} 
\end{center} 
\caption{Time evolution of overlap when model storing sequence B is obtained by (a) computer simulations ($N=100,000$) and (b) theoretical calculation and (c) represents the common external input given in eq.~(\ref{eq:patten_input}), where $\varepsilon=0.1$, $\sigma=0.1$, $b_2^t = 0.2$, and $c = 0.05$. } 
\label{fig:ovlp_pattern_input} 
\end{figure}

We discussed the model in which a common external input obeys the Gaussian distribution. We can, however, choose a certain common external input. We found empirically that a particular common external input caused a state transition. For instance, the common external input 
\begin{equation} 
\eta^t = \left\{
\begin{array}{rcl} 1 & , & $t \mbox{mod} 50 = 0$ \\ 0.6 & , & 0 < $t \mbox{mod} 50$ \, \le 3 \\ 0 & , & \mbox{otherwise} 
\label{eq:patten_input}\\ 
\end{array}\right. , 
\end{equation} 
makes the state transit along sequence B. To store sequence B, transition matrix $A$ becomes 
\begin{equation} 
A = \left( 
\begin{array}{cccccccc} 1 & 0 & 0 & 0 & 0 & 0 & 0 & \varepsilon\\ \varepsilon/3 & 1 & 0 & 0 & 0 & 0 & 0 & 0\\ \varepsilon/3 & 0 & 1 & 0 & 0 & 0 & 0 & 0\\ \varepsilon/3 & 0 & 0 & 1 & 0 & 0 & 0 & 0\\ 0 & \varepsilon & 0 & 0 & 1 & 0 & 0 & 0 \\ 0 & 0 & \varepsilon & 0 & 0 & 1 & 0 & 0 \\ 0 & 0 & 0 & \varepsilon & 0 & 0 & 1 & 0 \\ 0 & 0 & 0 & 0 & \varepsilon & \varepsilon & \varepsilon & 1 
\end{array} \right). 
\label{eq:Asammple2} 
\end{equation} 
Only memory pattern $\vec{\xi}^2$ is biased ($b_t^2 = 0.2, c = 0.05$). In this case, $m^{\mu}_{t+1}, \ \mu=1,2,...,8$ are given by eq.~(\ref{eqn:merf_bias_seq_a}) for limit $N \to \infty$. Figure~\ref{fig:ovlp_pattern_input} shows state transitions of model storing sequence B. It shows the time evolutions of the overlaps obtained by (a) computer simulation ($N=100,000$) and (b) theoretical calculation and (c) represents the particular common external input given in eq.~(\ref{eq:patten_input}). The parameters are $\varepsilon=0.1$, $\sigma=0.1$, $b_2^t = 0.2$, and $c = 0.05$. With the combination of such common external and bias inputs, state transitions with branching are controllable.

\section{Conclusion} 
We proposed an associative memory model that retrieved branching sequences using common external input and bias input. Since the model has sample dependence due to the common external input, we derived the macroscopic description as a probability density function of macroscopic variables. Then, we discussed the dynamics of the retrieval processes and effects of the bias input. The results obtained theoretically agree with those obtained by computer simulations. 

As a result, in the limit of $N \to \infty$, the model without the bias input transits to a mixed state at the branching point in the sequence. We found that the model with bias input, on the other hand, could transit to a target memory pattern. When this happens, the overlap between the bias input and the biased memory pattern, $b_t^{\mu}$, only needs to be large enough to break the symmetry of the mixed state. We also showed that state transition in branching sequences can be invoked by bias input and common external input, which takes a particular time structure, as shown in Fig.~\ref{fig:ovlp_pattern_input}. 

The transition mechanism that uses common external input is consistent 
with an experiment in songbirds, which suggests that the song control nucleus HVC 
receives common external input at the beginning of each song note\cite{Schmidt2003}. 
As mentioned above, Bengalese finches sing songs with finite-state 
automata and branching points\cite{Okanoya2004}. The state transition 
mechanism we have shown could play a role in the nervous system of Bengalese finches. 
Thus, our model has the advantage of being experimentally testable in an existing biological system. \\

\section*{Acknowledgments} 
This work was supported in part by a Grant-in-Aid for Scientific Research on Priority Areas No.~18020007 and No.~18079003, a Grant-in-Aid for Scientific Research (C) No.~16500093, a Grant-in-Aid for Young Scientists (B) No.~16700210, and the Director's fund of BSI, RIKEN.

\end{document}